\documentclass[10pt,pra,twocolumn,superscriptaddress,showpacs,floatfix]{revtex4}

\usepackage{graphicx}
\usepackage{amssymb}
\usepackage{times}
\usepackage{amsmath}
\usepackage{amsthm}

\input epsf.tex
\usepackage{graphicx}
\usepackage{amsthm}

\begin{document}

\title{Quantum Hacking on Quantum Key Distribution using Homodyne Detection}% Force line breaks with\\

\author{Jing-Zheng Huang}\affiliation
{Key Laboratory of Quantum Information, University of Science and Technology of China, Hefei, 230026, China}
\affiliation{Synergetic Innovation Center of Quantum Information $\&$ Quantum Physics, University of Science and Technology of China, Hefei, Anhui 230026, China}

\author{S\'{e}bastien Kunz-Jacques}\affiliation
{SeQureNet, 23 avenue d'Italie, 75013 Paris, France}

\author{Paul Jouguet}\affiliation
{SeQureNet, 23 avenue d'Italie, 75013 Paris, France}

\author{Christian Weedbrook}\affiliation
{Department of Physics, University of Toronto, Toronto, M5S 3G4, Canada}
\affiliation{QKD Corp., 112 College St., Toronto, M5G 1 L6, Canada}

\author{Zhen-Qiang Yin\footnote{yinzheqi@mail.ustc.edu.cn}}\affiliation
{Key Laboratory of Quantum Information, University of Science and Technology of China, Hefei, 230026, China}
\affiliation{Synergetic Innovation Center of Quantum Information $\&$ Quantum Physics, University of Science and Technology of China, Hefei, Anhui 230026, China}

\author{Shuang Wang\footnote{wshuang@ustc.edu.cn}}\affiliation
{Key Laboratory of Quantum Information, University of Science and Technology of China, Hefei, 230026, China}
\affiliation{Synergetic Innovation Center of Quantum Information $\&$ Quantum Physics, University of Science and Technology of China, Hefei, Anhui 230026, China}

\author{Wei Chen}\affiliation
{Key Laboratory of Quantum Information, University of Science and Technology of China, Hefei, 230026,  China}
\affiliation{Synergetic Innovation Center of Quantum Information $\&$ Quantum Physics, University of Science and Technology of China, Hefei, Anhui 230026, China}

\author{Guang-Can Guo}\affiliation
{Key Laboratory of Quantum Information, University of Science and Technology of China, Hefei, 230026,  China}
\affiliation{Synergetic Innovation Center of Quantum Information $\&$ Quantum Physics, University of Science and Technology of China, Hefei, Anhui 230026, China}

\author{Zheng-Fu Han}\affiliation
{Key Laboratory of Quantum Information, University of Science and Technology of China, Hefei, 230026, China}
\affiliation{Synergetic Innovation Center of Quantum Information $\&$ Quantum Physics, University of Science and Technology of China, Hefei, Anhui 230026, China}

\date{\today}

\begin{abstract}
Imperfect devices in commercial quantum key distribution systems open security loopholes that an eavesdropper may exploit. An example of one such imperfection is the wavelength dependent coupling ratio of the fiber beam splitter. Utilizing this loophole, the eavesdropper can vary the transmittances of the fiber beam splitter at the receiver's side by inserting lights with wavelengths different from what is normally used. Here, we propose a wavelength attack on a practical continuous-variable quantum key distribution system using homodyne detection. By inserting light pulses at different wavelengths, this attack allows the eavesdropper to bias the shot noise estimation even if it is done in real time. Based on experimental data, we discuss the feasibility of this attack and suggest a prevention scheme by improving the previously proposed countermeasures.
\end{abstract}

\maketitle

\section{Introduction}\label{Introduction}

Quantum key distribution~(QKD)~\cite{Gis2002,Sca2009} is a technology that provides a practical way to distribute a secret key between two distant parties using quantum physics and without making any assumptions on a potential eavesdropper's power. Such a level of theoretical security cannot be achieved using classical protocols. Recently, the study of the practical security of QKD systems~\cite{Sca2009} has attracted a lot of interest from the scientific community (see for example,~\cite{Hack1,Hack2,Hack3,Hack4,Hack5,Hack6,Hack7,Hack8,Li2011}). Indeed, deviations between the theoretical description of a QKD protocol and its implementation, open security loopholes that can be exploited by an eavesdropper. Demonstrations of partial or full eavesdropping against commercial discrete-variable QKD systems have been performed~\cite{Hack2,Hack3}. So far such hacking attacks were compiled on discrete variable systems as they were the only ones available at that time.

However, recently a commercial QKD system using continuous variables (CV)~\cite{Weedbrook2012}, that features secure distances~\cite{Jouguet1} comparable to commercial discrete-variable QKD systems, was released~\cite{sequrenet}. While the theoretical security of CV-QKD protocols has been established \cite{GC:prl06,NGA:prl06,Weedbrook2012,LGR:prl13}, the study of practical security of CV-QKD devices is far from sufficient (see for example, \cite{cvp1,cvp2,cvp3,cvp4,Jouguet3,Huang2013}). This is mostly due to the relative youth of the technology. Recent work includes the extension from discrete-variable QKD to CV-QKD of an attack (and solution) that exploits the wavelength dependency of fiber beam splitters~\cite{Li2011,Huang2013,Liang2013}. However, this attack was limited to the case where Bob performs heterodyne detection, i.e., he measures both quadratures of the electromagnetic field simultaneously~\cite{Weedbrook2004}. In this paper, we propose another wavelength dependency attack (along with a solution) but this time one that can be applied to a CV-QKD system using homodyne detection. Such a system also corresponds to those that are currently commercially available \cite{sequrenet}.

In Ref.~\cite{Jouguet2}, an attack targeting the local oscillator calibration routine of a CV-QKD system was proposed together with a family of countermeasures that consisted in measuring the shot noise in real time. We propose and provide experimental evidence of a wavelength attack targeting the real-time shot noise measurement procedure proposed in Ref.~\cite{Jouguet2}. By inserting light pulses at different wavelengths, this attack allows the eavesdropper to bias the shot noise estimation even if it is done in real time. Based on experimental evidence, we discuss the feasibility of this attack and suggest a prevention scheme by improving the previously proposed countermeasures.

In Sec.~\ref{background}, we first recall the basics of a CV-QKD scheme based on a Gaussian modulation of coherent states and homodyne detection. We present in detail how the relevant quantities, used to estimate the secret key rate of the protocol, are computed and tackle the problem of the shot noise evaluation procedure. Then, we give the principle of the attack proposed in \cite{Jouguet2} and the associated countermeasures. In Sec.~\ref{hack}, we explain how the wavelength dependency of the fiber beam splitter at the receiver's side can be exploited to bypass the real-time shot noise measurement countermeasure and detail the various steps of our attack. In Sec.~\ref{fea}, we study the practical feasibility of our scheme based on experimental values. Finally, we show in Sec.~\ref{countermeasure} how to improve the real-time shot noise measurement technique in order to detect our attack. The conclusion is given in Sec.~VI.

\section{Background}\label{background}

\subsection{CV-QKD using homodyne detection}\label{CVQKD}

\begin{figure}[!h]\center
\resizebox{9cm}{!}{
\includegraphics{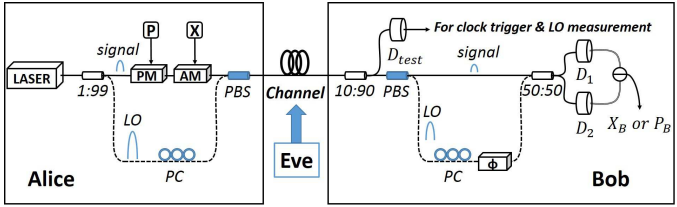}}
\caption{(Color online) A simplified schematic of the homodyne protocol scheme in Ref.~\cite{Jouguet1}. LO: local oscillator; PM: Phase modulator; AM: amplitude modulator; PC: polarization controller; PBS: polarization beam splitter; $D_1$, $D_2$ and $D_{test}$: detectors; $\phi$: phase modulator with $\phi = 0$ or $\pi/2$; (1:99), (10:90) and (50:50): beam splitters (reflectivity:transmittance).}\label{homodyne}
\end{figure}

A typical CV-QKD system using homodyne detection~\cite{Jouguet1} can be realized using the schematic given in Fig.~\ref{homodyne}. In this scheme, the weak signal and strong local oscillator are generated from the same coherent state pulse by a $1:99$ beam splitter. The signal is then modulated randomly following a Gaussian distribution with variance $V_A$ and zero mean in both quadratures, by using phase and amplitude modulators. The signal and local oscillator are separated in time and modulated into orthogonal polarizations using a polarization beam splitter before being inserted into the channel. When these pulses arrive at Bob's side, Bob randomly selects $\phi = 0$ or $\phi = \pi/2$ in order to measure either the $\hat{X}$ or $\hat{P}$ quadrature, respectively. After measuring, either direct or reverse error reconciliation (alternatively, postselection) protocols are performed in order to recover a common shared key. This is then followed by privacy amplification to reduce the eavesdropper's (Eve) knowledge to an arbitrary small amount~\cite{Sca2009,Weedbrook2012}.

\subsection{Homodyne detection and the local oscillator calibration attack}\label{homo}

Homodyne detection plays a key role in CV-QKD implementations. To illustrate the new wavelength attack scheme, let us first review the physical description of the homodyne detection. Note that a more detailed explanation can be found in Appendix~\ref{appendix 1}. Here we assume that both the signal and the local oscillator are coherent states. The signal state is denoted as $\alpha_s$ and the local oscillator is denoted as $\alpha_{LO}$. The specific quadrature of the signal is related to Bob's modulated phase $\phi$ and the substraction of the detector outcomes~\cite{cv-rev}. This can be expressed as
\begin{equation}
\begin{array}{lll} \label{xp}
{\hat{\delta}}{i} = \hat{i}_1 - \hat{i}_2 = \sqrt{\eta}\alpha_{LO}(X_{\phi} + \delta\hat{X}_{\phi}).
\end{array}
\end{equation}
Here $\hat{i}_1$ and $\hat{i}_2$ are the photocurrents recorded by detector 1 and detector 2, respectively; $\eta$ is the efficiency of the detectors; $X_{\phi} \equiv \alpha_s e^{-i\phi} + \alpha_s^* e^{i\phi}$ is the quadrature of the signal and $\delta\hat{X}_{\phi} \equiv \delta\hat{\alpha}_s e^{-i\phi} + \delta\hat{\alpha}_s^{\dag}e^{i\phi}$ is the quadrature of the vacuum state. When $\phi = 0$, $X_{0} \equiv X = \alpha_s + \alpha_s^*$ and when $\phi = \pi/2$, $X_{\pi/2} \equiv P = i(\alpha_s^* - \alpha_s)$.

A clock signal, which is generated by the local oscillator in a practical CV-QKD system (see Fig.~\ref{homodyne}), is necessary for maximizing the output of the homodyne detection. However, it opens a potential loophole for the eavesdropper.
In Ref.~\cite{Jouguet2}, the local oscillator calibration attack was proposed, in which
Eve modifies the shape of the local oscillator pulse in order to induce a delay to the clock trigger.
%\textbf{XXX Explain what the trigger does here.}
As a result, the homodyne detection outcome will drop down after such a delay due to the circuit design, which results in a decrease of the detection response slope, i.e., between the variance of the homodyne measurement and the local oscillator power. The value of the shot noise will be overestimated and consequently the excess noise present will be underestimated. Hence, Eve's presence will be underestimated.

\begin{figure}[!h]\center
\resizebox{8cm}{!}{
\includegraphics{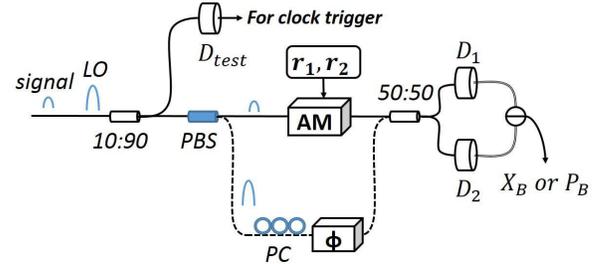}}
\caption{(Color online) Real-time shot noise measurement as proposed in Ref.~\cite{Jouguet2}. Here $r_1$ and $r_2$ denote the attenuation ratios applied on the signal path by the amplitude modulator. See text for more details.}\label{homo2}
\end{figure}

To prevent this attack, Bob can apply real-time shot noise measurements, which consists of two types of implementations~\cite{Jouguet2}. In this paper, we concentrate on the first one as shown in Fig.~\ref{homo2}. In this scheme, an amplitude modulator is added on the signal path. Bob randomly applies attenuation ratios $r_1 \approx 0$ and $r_2 \approx 1$ by the amplitude modulator to measure the shot noise level in real time. The measurement results will be directly used to estimate the shot noise in the data processing that follows.

\subsection{Fused biconical taper beam splitter}\label{fiber beam splitter}

Fiber beam splitters are one of the key components in an all-fiber QKD system. The most widely used technology in making fiber beam splitters is the so-called fused biconical taper technology~\cite{bs-1,bs-2}. As is described in Ref.~\cite{Li2011}, the coupling ratio of fused biconical taper beam splitter varies with the wavelength of the input light.
%
%The coupling ratio of FBT BS is given by\cite{bs-2}
%
%\begin{equation}
%\begin{array}{lll}\label{fiber beam splitter}
%R = F^2sin^2(\frac{C}{F}Z),
%\end{array}
%\end{equation}
%
%with
%
%\begin{equation}
%\begin{array}{lll}\label{fiber beam splitter}
%C = 0.021~\lambda^{2.5}/r^{3.5}, \\
%F = [1 + (\frac{234r^3}{\lambda^3})(\frac{\delta r}{r})^2]^{-1}, \\
%r = r_0$exp$(-\frac{L}{Z}).
%\end{array}
%\end{equation}
%
%where $\lambda$ is the wavelength,

For sufficing different requirements, there are three types of fused biconical taper beam splitters: the single wavelength type, the wavelength flatten type and the double wavelength type. Compared to the first two, the double wavelength type fused biconical taper beam splitter is more popular commercially because of its relatively stable performance in a wide wavelength range. Even so, it does not mean that it is totally wavelength independent. We experimentally tested two double wavelength type $10:90$ (reflection/transmittance) and a $50:50$  fused biconical taper beam splitter in our laboratory~\cite{thorlabs}.
%\textbf{XX these were actually tested in the lab??}
The relationship between their coupling ratios and wavelengths is shown in Table~I.
\begin{table}[h]
\label{thorlabs}
\begin{tabular}[c]{|c|c|c|c|c|c|c|}\hline
$\lambda$(nm) & 1270 & 1290 & 1310 & 1330 & 1350 & 1370 \\ \hline
$T$~ ($10:90$ BS) & 0.9050 & 0.9066 & 0.9020 & 0.8978 & 0.9014 & 0.8991 \\ \hline
$T$~ ($50:50$ BS) & 0.5327 & 0.5253 & 0.5144 & 0.5052 & 0.5011 & 0.4965 \\ \hline
$\lambda$(nm) & 1390 & 1410 & 1430 & 1450 & 1470 & 1490\\ \hline
$T$~ ($10:90$ BS) &0.8985 & 0.8938 & 0.8940 & 0.8985 & 0.8989 & 0.8985 \\ \hline
$T$~ ($50:50$ BS) &0.4931 & 0.4862 & 0.4902 & 0.4885 & 0.4908 & 0.4873 \\ \hline
$\lambda$(nm)  & 1510 & 1530 & 1550 & 1570 & 1590 & 1610 \\ \hline
$T$~ ($10:90$ BS) & 0.9012 & 0.8995 & 0.8956 & 0.9026 & 0.9022 & 0.9060 \\ \hline
$T$~ ($50:50$ BS) & 0.4954 & 0.4960 & 0.5012 & 0.5069 & 0.5155 & 0.5265 \\ \hline
\end{tabular}
\caption{The transmittance $T$ of Thorlabs~\cite{thorlabs} double wavelength type $10:90$ beam splitter and a $50:50$ beam splitter under different wavelengths $\lambda$~(nm).}
\end{table}

\section{Hacking homodyne detection systems}\label{hack}

In this section, a hacking scheme on a CV-QKD system using homodyne detection is proposed. Before introducing our scheme, two facts should be noted. First, in the improved CV-QKD scheme shown in Fig.~\ref{homo2}, Bob does not need to measure the intensity of the local oscillator because the shot noise level can be directly measured in real time. On the other hand, very low light intensity is enough to trigger the clock~\cite{Jouguet2}. In this case, Eve can hack the system by only utilizing the wavelength dependent character of the fused biconical taper beam splitter. A full wavelength attack scheme is proposed for this situation. Moreover, even if Bob monitors the local oscillator intensity, by combining the local oscillator calibration attack~\cite{Jouguet2} with the wavelength attack idea, Eve can still successfully achieve all of the secure key information without being discovered. Our attack scheme can be divided into two parts: Attack Part 1 and Attack Part 2.

\subsection{Attack Part 1}

In this attack Eve performs a full intercept-resend attack. For this purpose, she measures the information sent from Alice by performing heterodyne detection on both the signal and the local oscillator. After which she obtains two quadrature values $X_E$ and $P_E$. According to these measurement results, she prepares a new signal and local oscillator and sends them to Bob. In this stage, two strategies can be used.

\emph{Strategy A:} ~~Suppose that Bob does not monitor the local oscillator intensity.
Instead of preparing a signal state of amplitude $\alpha_E = \sqrt{\eta_{ch}}(X_E + iP_E)/2$ along with a local oscillator of amplitude $\alpha_{LO}$ as in the regular intercept-resend attack, Eve chooses a real number $N$ larger than 1 and prepares a signal state of amplitude $\sqrt{N}\alpha_E = \sqrt{N\eta_{ch}}(X_E + iP_E)/2$ along with a local oscillator of amplitude $\alpha_{LO}/\sqrt{N}$.
%
%Eve prepares the signal with quadratures $\sqrt{N\eta_{ch}} \hat{X}_E$ and $\sqrt{N\eta_{ch}} \hat{P}_E$, and the local oscillator with an amplitude of $\alpha_{LO}/\sqrt{N}$, where $N$ is a real number larger than 1 and $\eta_{ch}$ denotes the transmittance of channel.
The pulses are separated in time and orthogonal polarizations, as the original pulses were, and then sent onto Bob. In this strategy, Bob measures the quadratures with a variance of $\eta\eta_{ch}(V_A + 2N_0) + N_0/N + \eta\eta_{ch}\xi N_0 + v_{el}$ and a realistic shot noise of $N_0/N$, where $\xi$ is the excess noise in units of $N_0$, and $N_0 = \eta\alpha^2_{LO}$ (see Appendix~\ref{appendix 1} for details) is the shot noise variance without the attack.
%\textbf{XXXX - you can't reference an equation that hasn't happened yet!}
The excess noise Bob estimates is equal to $[2 + (1/N - 1)/\eta\eta_{ch} + \xi]N_0$~\cite{Jouguet2}. If he still uses $N_0$ as the shot noise unit, the excess noise he estimates can be made arbitrarily close to zero for certain channel efficiencies by choosing the proper $N$. For instance, by choosing typical values such as $\xi = 0.1$, $\eta = 0.5$~\cite{Jouguet2} and $N$ = 10, the excess noise estimated by Bob is $(2.1 - 1.8/\eta_{ch})N_0$. It reaches zero when $\eta_{ch} = 0.857$, corresponding to $0.67$~dB loss or about $3.35$~km of optical fiber link (with loss assumed to be $0.2$~dB/km). Thus entirely compromising the security of the protocol.

\emph{Strategy B:} ~~Suppose that Bob monitors the local oscillator intensity and its linear relation with the shot noise. And Eve performs the local oscillator calibration attack as proposed in Ref.~\cite{Jouguet2}. In this strategy, Eve controls the slope of the homodyne detection response by calibrating the trigger time. According to the analysis in~\cite{Jouguet2}, the excess noise estimated by Alice and Bob is close to zero when the realistic shot noise is reduced by $2/3$ of the original level and $\eta_{ch} = 0.5$.

Both of these strategies alone can not pass the protection test proposed in Ref.~\cite{Jouguet2} (Fig.~\ref{homo2}). Under this technique, Bob can easily monitor the shot noise level in real time, therefore he can modify the parameters immediately to fully protect against the above attacks. In order to not be discovered, Eve should take one more step to keep the counter-measurement results normal. For this purpose, the wavelength dependent character of fused biconical taper beam splitter is utilized to nullify the protection measurement in the second part of the scheme.

\subsection{Attack Part 2}

In this attack, Eve prepares and resends two extra coherent state pulses with wavelengths different from the typical communication wavelength of $1550$ nm. One of them is modulated the same polarization as the signal and the other with the local oscillator. So that when they reach Bob's side, one goes into the signal path and the other goes into the local oscillator path. Let us denote these pulses and also their intensities as $I^s$ and $I^{lo}$. Eve randomly chooses the wavelengths of $I^s$ and $I^{lo}$ from one of the following two sets:
\begin{equation}
\begin{array}{lll}\label{t}
\lambda_1^s &= 1410 nm, T_1^s = 0.4862,\\
\lambda_1^{lo} &= 1490 nm, T_1^{lo} = 0.4873;\\\\
\lambda_2^s &= 1310 nm, T_2^s = 0.5144,\\
\lambda_2^{lo} &= 1590 nm, T_2^{lo} = 0.5155,
\end{array}
\end{equation}
where $T^i_j$ $( i = s, lo; j = 1, 2 )$ denotes the transmittance of the $50:50$ fused biconical taper beam splitter corresponding to the different wavelengths (see Table~1). As the transmittances are deviated from $0.5$, an extra differential current proportional to the light intensity will appear in the final results.

When Bob applies strong attenuation ($r_1 \approx 0$) on the signal, the extra differential current is primarily contributed by $I^{lo}$. This extra contribution is equal to $(2T_1^{lo}-1)\eta_1^{lo} I_1^{lo} \equiv D^{lo}_1$ or $(2T_2^{lo}-1)\eta_2^{lo} I_2^{lo} \equiv D^{lo}_2$ plus shot noise (cf. Eq.~(\ref{photocurr2}) for details), where $\eta_i^{lo}$ denotes the detector efficiency corresponding to the different wavelengths. As this contribution should have zero statistical average and positive variance, Eve must ensure that $D^{lo}_1 = -D^{lo}_2$ and choose $I_1^{lo}$ and $I_2^{lo}$ with equal probability. In this case, the variance is approximately equal to $D_{lo}^2$. Therefore Eve should make $D_{lo}^2 = (1-1/N)N_0$ for Strategy A and $D_{lo}^2 = 1/3~N_0$ for Strategy B, in order to make the shot noise measurement results seem normal.

On the other hand, when Bob applies no attenuation ($r_2 \approx 1$) on the signal, the extra differential current comes from both $I^{lo}$ and $I^s$. Similarly, the differential current introduced by $I^s$ is $(1-2T_1^s)\eta_1^s I_1^s \equiv D^s_1$ or $(1-2T_2^s)\eta_2^s I_2^s \equiv D^s_2$ plus shot noise. Eve makes $D^s_1 = -D^s_2$ and chooses $I_1^s$ and $I_2^s$ with equal probability.
For convenience, we summarize the notations defined above as follows:
\begin{equation}
\begin{array}{lll} \label{notation}
D^s_1 \equiv (1-2T_1^s)\eta_1^s I_1^s,\\
D^{lo}_1 \equiv (2T_1^{lo}-1)\eta_1^{lo} I_1^{lo},\\\\
D^s_2 \equiv (1-2T_2^s)\eta_2^s I_2^s,\\
D^{lo}_2 \equiv (2T_2^{lo}-1)\eta_2^{lo} I_2^{lo}.
\end{array}
\end{equation}
By making $D^s_1 = -D^{lo}_1 = -D^s_2 = D^{lo}_2 \equiv D$, the contribution from $I^s$ will cancel the contribution from $I^{lo}$ except for a small amount of shot noise, which keeps the influence to the quadrature measurement results at an acceptable level. A more rigorous analysis taking the shot noises into account is described in Sec.~\ref{fea} and Appendix~\ref{appendix_2}.

%
%\begin{figure}[!h]\center
%\resizebox{9cm}{!}{
%\includegraphics{hacking1.eps}}
%\caption{The schematic of the wavelength attack.}\label{attack}
%\end{figure}
%

\section{Feasibility Analysis}\label{fea}

In this section, we analyze Bob's estimated excess noise under the two kinds of attacks proposed in Sec.~\ref{hack}. For simplicity, we take $\eta_i^j = \eta = 0.5$ ($i = 1, 2$; $j = s, lo$), $r_1 = 0.001$, $r_2 = 1$ and the intensity of the local oscillator $I_{LO} = 10^8$ (in units of photo-electron number). Let us  analyze the measurement outcomes corresponding to $I^{lo}$ and $I^s$.

First though, let us briefly review the method of estimating the excess noise in CV-QKD~\cite{Jouguet3}. By denoting $\hat{x}$ as the quadrature modulated by Alice ($\hat{X}_A$ or $\hat{P}_A$) and $\hat{y}$ as the quadrature measured by Bob ($\hat{X}_B$ or $\hat{P}_B$), we note that
%\textbf{XXX What's x and y? shouldn't it be X and P?? If so, change throughout.}
%
\begin{equation}
\begin{array}{lll} \label{stat1}
\langle \hat{x}^2\rangle = V_A N_0, \hspace{1mm} \langle \hat{x}\hat{y}\rangle = \sqrt{\eta\eta_{ch}}V_A N_0,\\
\langle \hat{y}^2\rangle = \eta\eta_{ch}(V_A + \xi)N_0 + N_0 + v_{el}.
\end{array}
\end{equation}
Here $\eta_{ch}$ is the channel transmittance, $V_A N_0$ is the modulation variance, $\xi$ is the excess noise, $N_0$ is the shot noise, $\eta$ is the efficiency of homodyne detector and $v_{el}$ is the electric noise (all expressed in their respective units). Among these parameters, $\eta$ and $v_{el}$ are pre-known as the system parameters, $N_0$ is estimated by the local oscillator intensity from $N_0 = \eta I_{LO}$, and the others are estimated from Alice and Bob's correlated variables $(x_i, y_i)_{i=1,...,m}$.
%\textbf{XXX - Isn't $\eta_{ch}$ only known through the correlated data as well??}
The excess noise can then be estimated as
\begin{equation}
\begin{array}{lll} \label{xi}
\xi = (\langle \hat{y}^2\rangle - \eta\eta_{ch}V_A N_0 - N_0 - v_{el})/\eta\eta_{ch}N_0.
\end{array}
\end{equation}

In a later protection scheme given in Ref.~\cite{Jouguet2}, two attenuation ratios, $r_1$ and $r_2$, are introduced on the signal path. Typically we set $r_1 = 0.001$ for shot noise estimation and $r_2 = 1$ for quadrature measurements. The variance of $\hat{y}$ should be expressed as
\begin{equation}
\begin{array}{lll} \label{y}
\langle\hat{y}^2\rangle_1 &\equiv V_{s1} = r_1\eta\eta_{ch}(V_A + \xi)N_0 + N_0 + v_{el}\\
\langle\hat{y}^2\rangle_2 &\equiv V_{s2} = r_2\eta\eta_{ch}(V_A + \xi)N_0 + N_0 + v_{el}.
\end{array}
\end{equation}
%
%Here we have assumed that the excess noise is proportional to the total efficiency on the signal.
We can then estimate the parameters by
%\textbf{XXXX - Below and throughout please check whether operators should and should not be put on variables!}
%
\begin{equation}
\begin{array}{lll} \label{para}
\tilde{N}_0 &= \frac{r_2V_{s1}-r_1V_{s2}}{r_2-r_1} - v_{el},\\
\tilde{\xi} &= [\frac{V_{s2}-V_{s1}}{(r_2-r_1)\eta\eta_{ch}} - V_A \tilde{N}_0]/\tilde{N}_0.
\end{array}
\end{equation}
From now on, we denote $N_0 \equiv \eta I_{LO}$ as a constant value (that is, the shot noise value when the system runs normally) and $\tilde{N}_0$ and $\tilde{\xi}$ as the estimation values. Let us now analyze how large $\tilde{N}_0$ and $\tilde{\xi}$
%\textbf{XXX - what is 'it'?}
could be under the two different attack strategies.

\emph{Strategy A:}~~The differential current $\hat{\delta i}_{tot}$ at the output of the homodyne detection can be considered as the summation of
%\textbf{XXXX - throughout the paper put the hat on the delta and not on the i.}
$\hat{\delta i}_{part1}$ and $\hat{\delta i}_{part2}$, which present the contributions from Part~1 and Part~2 of our attack scheme respectively. That is, $\hat{\delta i}_{tot,i} = \hat{\delta i}_{part1,i} + \hat{\delta i}_{part2,i}$, where the index $i = 1, 2$ denotes that Bob applies the attenuation ratio $r_i$. In Strategy A, $\hat{\delta i}_{part,i}$ can be obtained (cf. Eq.~(\ref{xpapp})) by taking $\hat{X}_{\phi} = \sqrt{r_i\eta_{ch}N}(X_A + \delta\hat{X}_A + \delta\hat{X}_E)$, and its variance can then be computed as
\begin{equation}
\begin{array}{lll} \label{vsa-1}
V^A_{part1,i} &= \langle(\hat{\delta i}_{part1,i})^2\rangle - \langle\hat{\delta i}_{part1,i}\rangle^2\\
          &= \eta\frac{\alpha^2_{LO}}{N}[r_i\eta\eta_{ch} N(V_A +2) + 1] + r_i\eta\eta_{ch}\xi N_0 + v_{el}\\
          &= r_i\eta\eta_{ch}(V_A + 2 + \xi)N_0 + \frac{N_0}{N} + v_{el}.
\end{array}
\end{equation}
For $\hat{\delta i}_{part2,i}$, we derive its variance in Appendix~\ref{appendix_2} (cf. Eq.~(\ref{vpart2i})) as follows
\begin{equation}
\begin{array}{lll} \label{vpart22}
V_{part2,i} = (1 - r_i)^2D^2 + (35.81 + 35.47r_i^2)D.
\end{array}
\end{equation}
Thus the total variance is given by
\begin{equation}
\begin{array}{lll} \label{vsa}
V^A_{s,i} &= \langle(\hat{\delta i}_{tot})^2\rangle - \langle\hat{\delta i}_{tot}\rangle^2\\
        &= V^A_{part1,i} + V_{part2,i}\\
        &= r_i\eta\eta_{ch}(V_A + 2 + \xi)N_0 + \frac{N_0}{N} + v_{el}\\
        &~~ + (1 - r_i)^2D^2 + (35.81 + 35.47r_i^2)D.
\end{array}
\end{equation}
We can now get the estimations about the shot noise level and excess noise under Strategy A to be
\begin{equation}
\begin{array}{lll} \label{para-a}
\tilde{N}_0 &= N_0/N + (1 - r_1r_2)D^2 + (35.81 - 35.47r_1r_2)D,\\
\tilde{\xi}_A &= [(2 + \xi)N_0 + (r_1 + r_2 - 2)D^2/\eta\eta_{ch} \\
            &~~+ 35.47(r_1+r_2)D]/\tilde{N}_0.
\end{array}
\end{equation}
By choosing proper intensities $I^s$, $I^{lo}$ and $N$, Eve can make $\tilde{N}_0 = N_0$ and $\tilde{\xi}_A$ arbitrary close to zero. For this purpose, we take $\eta_{ch} = 0.9$, for example. Assume $\xi = 0.1$, simple calculations show that Eve can choose $N = 20.9$, $I^s_1 = 5 \times10^5$, $I^{lo}_1 = 5.4 \times 10^5$, $I^s_2 = 4.8 \times 10^5$ and $I^{lo}_2 = 4.4 \times10^5$, which are $3$ orders of magnitude smaller than $I_{LO}$.

\emph{Strategy B:}~~As long as Eve can change the slope of the homodyne detection response by calibrating the trigger time, the excess noise $\tilde{\xi}_B$ will be close to zero. Let us assume the realistic shot noise is $\gamma N_0$. It is easy to derive that
\begin{equation}
\begin{array}{lll} \label{vsb-1}
V^B_{part1,i} = \gamma[r_i\eta\eta'_{ch}(V_A + 2 + \xi) + 1]N_0 + v_{el},
\end{array}
\end{equation}
and $V_{part2,i}$ is the same as in Strategy A. Here $\gamma$ and $\eta'_{ch}$ are parameters chosen by Eve, and she should make $\gamma\eta'_{ch} = \eta_{ch}$ in order to keep the estimated parameters normal.
Therefore
\begin{equation}
\begin{array}{lll} \label{vsb}
V^B_{s,i} &= \gamma[r_i\eta\eta'_{ch}(V_A + 2 + \xi) + 1]N_0 + v_{el}\\
              &~~ + (1 - r_i)^2D^2 + (35.81 + 35.47r_i^2)D.
\end{array}
\end{equation}
The shot noise level and excess noise under Attack 2 can then be computed to give
\begin{equation}
\begin{array}{lll} \label{para-b}
\tilde{N}_0 &= \gamma N_0 + (1 - r_1r_2)D^2 + (35.81 - 35.47r_1r_2)D,\\
\tilde{\xi}_B &= [(2 + \xi)N_0 + V_A(N_0 - \tilde{N}_0)\\
            &~~ + (r_1 + r_2 - 2)D^2/\eta\eta_{ch} + 35.47(r_1+r_2)D]/\tilde{N}_0.
\end{array}
\end{equation}
By choosing proper intensities $I^s$, $I^{lo}$ and $N$, Eve can make $\tilde{N}_0 = N_0$ and $\tilde{\xi}_B$ arbitrarily close to zero. Let us take $\eta_{ch} = 0.5$ and $\xi = 0.1$, for example. Again a simple calculation shows that, by choosing $\gamma = 0.47$, $I^s_1 = 3.72 \times 10^5$, $I^{lo}_1 = 4.04 \times 10^5$, $I^s_2 = 3.56 \times 10^5$ and $I^{lo}_2 = 3.31 \times 10^5$, we again get about $3$ orders of magnitude smaller than $I_{LO}$ as in Strategy A.

Finally, we note that the intensities of the pulses in Part 2 will affect the local oscillator intensity measurement. This effect is small due to the low strength of the pulses in Part 2, and Eve can fully compensate it by decreasing the local oscillator intensity in Part 1 and carefully calibrating the trigger time.

\section{Countermeasure: improvement of the real-time shot noise measurement technique}
\label{countermeasure}

\newcommand{\var}{\operatorname{Var}}
\newcommand{\cov}{\operatorname{Cov}}

In the former proposed scheme in the real-time shot noise measurement regime~\cite{Jouguet2}, only two attenuation ratios $r_1 \approx 0$ and $r_2 = 1$ are applied on the signal path, and we have already shown that this is not enough to detect the wavelength attack.
In fact, in that case, according to Eqs.~(\ref{vsa}) and (\ref{vsb}), the total noise $N(r)$ can be written as a second-order polynomial of the attenuation ratio $r$:
\begin{equation}
\label{vsi}
\begin{split}
N(r) =
r^2 \var(X_s) &+ r (\eta\eta_{ch}\xi - 2 \cov(X_s, X_{lo}))\\ &+ (N_0 + v_{el} + \var(X_{lo})),
\end{split}
\end{equation}
where $X_s$ ($X_{lo}$) is the signal on the detection caused by the attack signal going through the signal path (local oscillator path). The shot noise measurement procedure in Ref.~\cite{Jouguet2} assumes that $V(r)$ is a linear function of $r$ and this is why it is defeated by the wavelength attack, which uses the term $\cov(X_s, X_{lo})$ to compensate for the terms $\var(X_s)$ and $\var(X_{lo})$ when $r=1$.

The countermeasure can be modified to thwart the wavelength attack by allowing Bob to use a third attenuation ratio, thereby observing $N(r)$ for three values of $r$. This way the three coefficients $a,b,c$ of the polynomial $V(r) = a r^2 + br+c$ can be obtained. The coefficient $a$ in front of $r^2$ should be 0 in an ideal setting. To avoid the wavelength attack, it is enough that Alice and Bob ensure that $a \ll c$. Indeed, in that case,
$$\var(X_s) \ll N_0 + \var(X_{lo}),$$ hence, $$\var(X_s) \ll \var(X_{lo}),$$ (since $\var(X_{lo})$ is not small compared to $N_0$). As a result,
$\cov(X_s, X_{lo}) \leq \sqrt{\var(X_s) \var(X_{lo})} \ll \var(X_{lo})$, and it is not possible anymore to compensate $\var(X_{lo})$ with $\cov(X_s, X_{lo})$.

For instance, Bob can randomly apply attenuation ratios $r_1 = 1$, $r_2 = 0.5$ and $r_3 = 0.001$ to the amplitude modulator, with probabilities of $90\%$, $5\%$ and $5\%$ respectively. As has been pointed out in Ref.~\cite{Jouguet2}, this countermeasure has an impact on the overall key rate since some pulses are attenuated. In our example, assuming that $10\%$ of the pulses that are attenuated are discarded, the final key rate is the same as in Ref.~\cite{Jouguet2}.

It is worth noting that applying randomly several attenuation ratios on Bob's side allows us to check the transmittance linearity with respect to the attenuation ratio in the same way as we do for the noise. This allows us for instance to defeat saturation attacks \cite{Qin2013} that rely on non-linearities of the detection apparatus. Therefore this countermeasure defeats all currently known attacks on the detection apparatus of Gaussian CVQKD, and is expected to constitute a strong defense against variants of these attacks.

In addition to the procedure above, physical countermeasures such as adding wavelength filters before detection (to ensure that the wavelengths used for the attacks are close to the system wavelength, which forces the attacker to use high-power signals), and a monitoring of the local oscillator intensity (to detect these high-power signals) are also suggested.

\section{Conclusion}\label{conclusion}

In conclusion, we proposed two strategies to realize a wavelength attack targeting a practical CV-QKD system using homodyne detection. By inserting light pulses at different wavelengths, with intensities lower than the local oscillator light by three orders of magnitude, Eve can bias the shot noise and the excess noise estimated by Alice and Bob. In other words, Eve can tap all of the secure key information without being discovered. The real-time shot noise measurement scheme as proposed in Ref.~\cite{Jouguet2} cannot detect this type of attack. However, it can be improved by using three attenuation ratios to successfully fix this security loophole. Moreover, other physical countermeasures, such as adding additional wavelength filters and monitoring the local oscillator intensity, are also suggested.

%Finally, we note that this improved countermeasure regime prevents not only the wavelength attack, but probably also the saturation attack proposed in \cite{Qin2013}.

\acknowledgments

We thank Xiao-Tian Song and Yun-Guang Han for providing the test data. This work was supported by the National Basic Research Program of China (Grants No. 2011CBA00200 and No. 2011CB921200), National Natural Science Foundation of China (Grants No. 60921091 and No. 61101137).
%NSERC, and QuantumWorks.
P. J. and S. K.-J. acknowledge support from the French National Research Agency, through the HIPERCOM (2011-CHRI-006) project, by the DIRECCTE Ile-de-France through the QVPN (FEDER-41402) project, and by the European Union through the Q-CERT (FP7-PEOPLE-2009-IAPP) project. C. W. acknowledges support from NSERC.

%%%%%%%%%%%%%%%%%%%%%%%%%%%%%%%%%%%%%%%%%%%%%%%%%%%%%%%%%%%%%%%%%%%%%%%%%%%%%%%%%%%%%%%%%%

\appendix
\section{Mathematical Model of Homodyne Detection}\label{appendix 1}

When a signal, described by the annihilation operator $\hat{a}$, is inserted to a photodetector with an efficiency of $\eta$, the measured annihilation operator becomes $\hat{b} = \sqrt{\eta}\hat{a} + \sqrt{1-\eta}\hat{a}_v$, where $\hat{a}_v$ denotes the vacuum mode.
The input photons are converted to an electric current with strength $\hat{i} = q\hat{b}^{\dag}\hat{b}$, where $q$ is a constant amplification factor and $\hat{i}$ represents the number of electrons. Without loss of generality, we set $q = 1$ for simplicity.
\begin{figure}[!h]\center
\resizebox{6cm}{!}{
\includegraphics{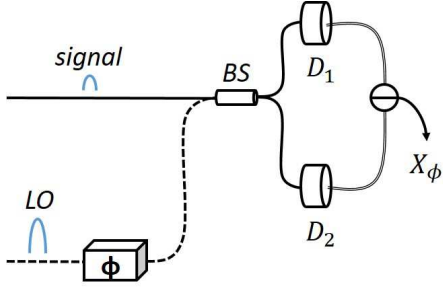}}
\caption{(Color online) Schematic of the basics of homodyne detection. See text for details.}\label{homodet}
\end{figure}
The model of homodyne detection is shown in Fig.~\ref{homodet}. Here we assume that the transmittance of the beam splitter is $T$, the photocurrents recorded by the photodetectors can be written as follows (where the electric noise is not considered)
\begin{equation}\label{photocurr}
\begin{array}{lll}
\hat{i}_1 &= [\sqrt{\eta}(e^{-i\phi}\sqrt{T}\hat{a}_{LO}^{\dag} - \sqrt{1-T}\hat{a}_s^{\dag}) + \sqrt{1-\eta}\hat{a}^{\dag}_{v1}]\times\\
          &[\sqrt{\eta}(e^{i\phi}\sqrt{T}\hat{a}_{LO} - \sqrt{1-T}\hat{a}_s) + \sqrt{1-\eta}\hat{a}_{v1}],\\
\hat{i}_2 &= [\sqrt{\eta}(e^{-i\phi}\sqrt{1-T}\hat{a}_{LO}^{\dag} + \sqrt{T}\hat{a}_s^{\dag}) + \sqrt{1-\eta}\hat{a}^{\dag}_{v2}]\times\\
          &[\sqrt{\eta}(e^{i\phi}\sqrt{1-T}\hat{a}_{LO} + \sqrt{T}\hat{a}_s) + \sqrt{1-\eta}\hat{a}_{v2}],\\
\end{array}
\end{equation}
where $\phi \in \{0, \pi/2\}$ is switchable and controlled by Bob.
We note that $\hat{a}_s$ can be linearized and written as $\alpha_s + \delta\hat{a}_s$ and $\hat{a}_{LO}$ can be written as $\alpha_{LO} + \delta\hat{a}_{LO}$, where $\delta\hat{a}_s$ and $\delta\hat{a}_{LO}$ can be considered as the annihilation operators of the vacuum state~\cite{book1}. For simplicity, let us assume that $\alpha_{LO}$ is a real number. To derive the quadratures $\hat{X}$ and $\hat{P}$, the difference of the two photocurrents should be measured:
\begin{equation}
\begin{array}{lll} \label{photocurr2}
\hat{\delta i} &= \hat{i}_2 - \hat{i}_1\\
              &= \eta[(1-2T)\hat{a}^{\dag}_{LO}\hat{a}_{LO}\\
              &~~+ 2\sqrt{T(1-T)}(e^{-i\phi}\hat{a}^{\dag}_{LO}\hat{a}_s + e^{i\phi}\hat{a}_{LO}\hat{a}^{\dag}_s) + (2T-1)\hat{a}^{\dag}_s\hat{a}_s] \\
              &~~+ \sqrt{\eta(1-\eta)}[\sqrt{T}(e^{-i\phi}\hat{a}^{\dag}_{LO}\hat{a}_{v1} + e^{i\phi}\hat{a}_{LO}\hat{a}^{\dag}_{v1})\\
              &~~+ \sqrt{1-T}(e^{-i\phi}\hat{a}^{\dag}_{LO}\hat{a}_{v2} + e^{i\phi}\hat{a}_{LO}\hat{a}^{\dag}_{v2})]\\
              &\simeq \sqrt{\eta}\alpha_{LO}[(2T-1)\sqrt{\eta}(\alpha_{LO} + \delta\hat{X}_{LO}) \\
              &~~~~~~~~~~~~~~~~~~~+ 2\sqrt{T(1-T)}\sqrt{\eta}(X_{\phi} + \delta\hat{X}_{\phi})\\
              &~~~~~~~~~~~~~~~~~~~+ \sqrt{(1-\eta)}(\sqrt{T}\hat{X}_{v1} + \sqrt{1-T}\hat{X}_{v2})],
\end{array}
\end{equation}
where $X_{\phi} \equiv \alpha_s e^{-i\phi} + \alpha_s^* e^{i\phi}$, $\delta\hat{X}_{LO} \equiv \delta\hat{a}_{LO} + \delta\hat{a}_{LO}^{\dag}$ and $\delta\hat{X}_{\phi} \equiv \delta\hat{a}_s e^{-i\phi} + \delta\hat{a}_s^{\dag}e^{i\phi}$. Here $\delta\hat{X}_{\phi}$, $\hat{X}_{v1}$ and $\hat{X}_{v2}$ are irrelevant vacuum states with a normalized variance of 1. When $\phi = 0$ (or $\pi/2$) we recover the quadratures $\hat{X}$ (or $\hat{P}$) from $X_{\phi}$, respectively. The approximation comes from the fact that $\alpha_s \ll \alpha_{LO}$. By setting $T = 0.5$, we have
\begin{equation}
\begin{array}{lll} \label{xpapp}
\hat{\delta i} &= \sqrt{\eta}\alpha_{LO}(\sqrt{\eta}X_{\phi} + \sqrt{\eta}\delta\hat{X}_{\phi} + \sqrt{1-\eta}\frac{\hat{X}_{v1} + \hat{X}_{v2}}{\sqrt{2}})\\
              &= \sqrt{\eta}\alpha_{LO}\hat{X}^{\phi}_{Bob}
\end{array}
\end{equation}
which is called balanced homodyne detection. Bob can calculate the variance of $\hat{X}^{\phi}_{Bob}$ from $\hat{\delta i}$ by $V^{\phi}_B = \langle\hat{\delta i}^2\rangle = \eta\alpha^2_{LO}\langle(\hat{X}^{\phi}_{Bob})^2\rangle = \eta V_{\phi}N_0 + N_0$, where $N_0 \equiv \eta\alpha^2_{LO}$ is used as the shot noise unit and $V_{\phi} \equiv \langle X^2_{\phi}\rangle$ is the variance of $X_{\phi}$ in shot noise units.

In general, we can calculate the variance of $\hat{\delta i}$ as
\begin{equation}
\begin{array}{lll} \label{vi}
\langle\hat{\delta i}^2\rangle &= \eta^2\alpha^4_{LO}(2T-1)^2 + \eta^2\alpha^2_{LO}(2T-1)^2\langle\delta\hat{X}^2_{LO}\rangle\\
                              &~~+ 4\eta^2\alpha^2_{LO}T(1-T)(\langle X^2_{\phi}\rangle + \langle\delta\hat{X}^2_{\phi}\rangle)\\
                              &~~+ \eta(1-\eta)\alpha^2_{LO}[T\langle\hat{X}^2_{v1}\rangle + (1-T)\langle\hat{X}^2_{v2}\rangle]\\
                              &= \eta^2\alpha^4_{LO}(2T-1)^2 + \eta\alpha^2_{LO}[1-\eta+\eta(2T-1)^2]\\
                              &~~+ 4\eta^2\alpha^2_{LO}T(1-T)(\langle X^2_{\phi}\rangle + 1)
\end{array}
\end{equation}
Finally, the differential current introduced by excess noise and electric noise should be added~\cite{chi2011}. Hence, the total output current is $\hat{\delta i}_{tot} = \hat{\delta i} + \hat{\delta i}_{\xi} + \hat{\delta i}_{el}$ with a variance of $\langle\hat{\delta i}^2_{tot}\rangle = \langle\hat{\delta i}^2\rangle + \langle\hat{\delta i}^2_{\xi}\rangle + \langle\hat{\delta i}^2_{el}\rangle = \langle\hat{\delta i}^2\rangle + \eta\xi N_0 + v_{el}$.

\section{Analysis of the Attack Part 2 Excess Noise}\label{appendix_2}

According to Eq.~(\ref{photocurr2}), the differential currents introduced by $I^{lo}$ and $I^s$ are
\begin{equation}
\begin{array}{lll} \label{fshot}
\hat{\delta i}_{lo} &= \eta(2T^{lo}_j-1)I^{lo}_j + \sqrt{\eta I^{lo}_j}\left[\left(2T^{lo}_j-1\right)\sqrt{\eta}\delta\hat{X}^{lo}_j \right.\\
              &~~~~~~~~~~~~~~~~~~~+ 2\sqrt{T^{lo}_j(1-T^{lo}_j)}\sqrt{\eta}\delta\hat{X}_{\phi 1} \\
              &~~~~~~~~~~~~~~~~~~~+ \sqrt{(1-\eta)}\left.\left(\sqrt{T^{lo}_j}\hat{X}_{v1} + \sqrt{1-T^{lo}_j}\hat{X}_{v2}\right)\right]\\
              &= D^{lo}_j + \hat{S}^{lo}_j,\\
\hat{\delta i}_s &= \eta(1-2T^s_j)I^s_j + \sqrt{\eta I^s_j}\left[\left(1-2T^s_j\right)\sqrt{\eta}\delta\hat{X}^s_j \right.\\
              &~~~~~~~~~~~~~~~~~~~- 2\sqrt{T^s_j(1-T^s_j)}\sqrt{\eta}\delta\hat{X}_{\phi 2} \\
              &~~~~~~~~~~~~~~~~~~~- \sqrt{(1-\eta)}\left.\left(\sqrt{T^s_j}\hat{X}_{v3} + \sqrt{1-T^s_j}\hat{X}_{v4}\right)\right]\\
              &= D^s_j + \hat{S}^s_j,
\end{array}
\end{equation}
where $\hat{S}^{lo}_j$ and $\hat{S}^s_j$ denote the shot noise terms and $j = 1$ or $2$ corresponds to the random choices.
Let us denote the summation of these two contributions for attenuation ratio $r_i$ by $\hat{\delta i}_{part2,i} = \hat{\delta i}_{lo} + r_i\hat{\delta i}_s$ ($i = 1, 2$). The variance of $\hat{\delta i}_{part2,i}$ can be computed by
\begin{equation}
\begin{array}{lll}\label{vpart2}
V_{part2,i} &= \langle\hat{\delta i}_{part2,i}^2\rangle - \langle\hat{\delta i}_{part2,i}\rangle^2\\
         &= \langle(D^{lo}_j + r_iD^s_j + \hat{S}^{lo}_j + r_i\hat{S}^s_j)^2\rangle - 0\\
         &= \langle(D^{lo}_j + r_iD^s_j)^2\rangle + \langle(\hat{S}^{lo}_j)^2\rangle + \langle(r_i\hat{S}^s_j)^2\rangle\\
         &= \langle(-D^s_j + r_iD^s_j)^2\rangle\\
         &~~+\eta\langle I^{lo}_j[\eta(2T^{lo}_j-1)^2 + 4\eta T^{lo}_j(1-T^{lo}_j) + 1 - \eta]\rangle\\
         &~~+ \eta r_i^2\langle I^s_j[\eta(1-2T^s_j)^2 + 4\eta T^s_j(1-T^s_j) + 1 - \eta]\rangle\\
         &= (1 - r_i)^2D^2 + \eta\langle I^{lo}_j\rangle + \eta r_i^2\langle I^s_j\rangle,\\
\end{array}
\end{equation}
where we have used the conditions proposed in Sec.~\ref{hack}, which can be rearranged as $D_1^s = -D_2^s = -D_1^{lo} = D_2^{lo} = D$. $I^n_j$ can be expressed in terms of $D^n_j$ ($n = s, lo$; $j = 1, 2$) from their definitions (cf. Eq.~(\ref{notation})). That is,
\begin{equation}
\begin{array}{lll} \label{imean1}
\eta\langle I^{lo}_j\rangle    &= \frac{\eta}{2}I^{lo}_1 + \frac{\eta}{2}I^{lo}_2\\
                               &= \frac{\eta}{2}\frac{D^{lo}_1}{\eta(2T^{lo}_1 - 1)} + \frac{\eta}{2}\frac{D^{lo}_2}{\eta(2T^{lo}_2 - 1)}\\
                               &= 35.81D,
\end{array}
\end{equation}
\begin{equation}
\begin{array}{lll} \label{imean2}
\eta r_i^2\langle I^s_j\rangle &= r_i^2(\frac{\eta}{2}I^s_1 + \frac{\eta}{2}I^s_2)\\
                               &= r_i^2(\frac{\eta}{2}\frac{D^s_1}{\eta(1 - 2T^s_1)} + \frac{\eta}{2}\frac{D^s_2}{\eta(1 - 2T^s_2)})\\
                               &= 35.47r_i^2D.
\end{array}
\end{equation}
For getting the real numbers, we have substituted $T_1^s$, $T_1^{lo}$, $T_2^s$ and $T_2^{lo}$ by their values according to Eq.~(\ref{t}).
Therefore, we finally get
\begin{equation}
\begin{array}{lll} \label{vpart2i}
V_{part2,i} = (1 - r_i)^2D^2 + (35.81 + 35.47r_i^2)D.
\end{array}
\end{equation}

\end{document}